\documentclass[twocolumn,showpacs,preprintnumbers,amsmath,amssymb,pra,floatfix]{revtex4}

\usepackage{graphicx} 
\usepackage{dcolumn} 
\usepackage{bm} 
\usepackage{longtable}

\begin{document}

\preprint{APS/123-QED}

\title{Experimental and theoretical lifetimes and transition probabilities in Sb I }

\author{Henrik Hartman, Hampus Nilsson}
\affiliation{Department of Astronomy and Theoretical Physics, Lund University, Box 43, SE-221 00 Lund, Sweden\\    \texttt{Henrik.Hartman; Hampus.Nilsson@astro.lu.se}}

\author{Lars Engstr\"om, Hans Lundberg}
\affiliation{Department of Physics, Lund University, Box 118, SE-221 00 Lund, Sweden\\
\texttt{Lars.Engstrom; Hans.Lundberg@fysik.lth.se}
}

\author{Patrick Palmeri, Pascal Quinet$^*$, and \'Emile Bi\'emont}
 \altaffiliation[Also at ]{IPNAS, Universit\'e de Li\`ege, B-4000, Li\`ege, Belgium.}
\affiliation{Astrophysique et Spectroscopie, Universit\'e de Mons-UMONS, Place du Parc, 20, B-7000 Mons, Belgium\\ 
\texttt{patrick.palmeri@umons.ac.be; Pascal.Quinet;  E.Biemont@ulg.ac.be}
}

\date{\today} 

\begin{abstract}

We present experimental atomic lifetimes for 12 levels in Sb I, out of which seven are reported for the first time. The levels belong to the 5p$^2$($^3$P)6s $^{2}$P, $^{4}$P and 5p$^2$($^3$P)5d $^{4}$P, $^{4}$F and $^{2}$F terms. The lifetimes were measured using time-resolved laser-induced fluorescence. In addition, we report new calculations of transition probabilities in Sb I using a Multiconfigurational Dirac-Hartree-Fock method. The physical model being tested through comparisons between theoretical and experimental lifetimes for 5d and 6s levels. The lifetimes of the 5d $^4$F$_{3/2, 5/2, 7/2}$ levels (19.5, 7.8 and 54 ns, respectively) depend strongly on the $J$-value. This is explained by different degrees of level mixing for the different levels in the $^4$F term.

\end{abstract}


\maketitle

\section{\label{sec:level1}Introduction}

As many other spectra, Sb~I was first investigated by Meggers \& Humphreys \cite{MH42}. Since then the analysis has been extended and revised by Mazzoni \& Joshi \cite{MJ79}, Joshi  {\it et~al.} \cite{JSV84}, Zaidi {\it et~al.} \cite{ZMB84}, Beigang \& Wynne \cite{BW86} and Voss {\it et~al.} \cite{VWBW86}. The most recent and complete work is that of Hassini  {\it et~al.} \cite{HARVW88} who reported 138 levels derived from 617 spectral lines measured in the range 2536 to 24786 cm$^{-1}$,  using Fourier transform spectroscopy.  Of the reported levels 32 were previously unknown and several previous levels assignments were revised. Furthermore, Hassini  {\it et~al.} \cite{HARVW88} reported hyperfine splitting constants for more than 75 \% of the levels.

Belin {\it et~al.} \cite{BGHR74} measured radiative lifetimes of the two levels ($^3$P)6s $^4$P$_{3/2}$ and $^4$P$_{5/2}$ using the Hanle-effect method. Andersen {\it et~al.} \cite{AWS74} measured six levels (($^3$P)6s $^4$P$_{1/2,3/2,5/2}$, $^2$P$_{3/2}$, and ($^1$D)6s $^2$D$_{3/2,5/2}$) using the beam-foil method. Furthermore, 14 lifetimes measured with beam foil, level-crossing and multichannel delayed-coincidence detection were reported by Osherovich \& Tezikov \cite{OT77} and Tezikov \cite{T78} for levels in the 6s, 7s, and 8s configurations. 

Transition probabilities and lifetimes in neutral antimony are still very scarce.  In astrophysics, this explains e.g. why the most recent compilation of solar abundances \cite{GASS} does not provide any result for the photospheric abundance of this element while the meteoritic result is well establised. This is partly due to the fact that the lines identified as Sb I in the solar  spectrum \cite{MMH} are severely blended but also to the fact that reliable oscillator strengths are still lacking for these transitions.

In the present paper, we report experimental lifetimes for 12 Sb~I levels measured with time-resolved laser-induced fluorescence (TR-LIF) spectroscopy. Seven of the levels are reported for the first time. In addition, we report calculated lifetimes and transition probabilities using a Multiconfigurational Dirac-Hartree-Fock (MCDHF) method. The accuracy of the calculations is assessed through the good agreement observed between theoretical and available experimental lifetimes and branching fractions. 

\section{The Sb I term system}

The spectrum of neutral antimony (Sb~I) is homologous with that of N~I, with the lowest configuration $n$s$^2$$n$p$^3$ (in Sb~I, $n=5$). Due to the Pauli exclusion principle, this configuration contains only three terms: $^4$S$^{\circ}$, $^2$D$^{\circ}$ and $^2$P$^{\circ}$, out of which $^4$S$^{\circ}$ is the ground state. The system, with one excited p-electron, can be written as $5$s$^2$$5$p$^2$ $nl$, and has the three parent terms, $^3$P$^{\circ}$, $^1$D$^{\circ}$ and $^1$S$^{\circ}$, which are the lowest terms in Sb~II. The lowest even configuration is ($^3$P)6s containing a $^4$P and a $^2$P term. There are no experimentally known levels belonging to the doubly excited system 5s$^2$5p $nln'l'$, or to the system where one or both of the 5s electrons are excited.   

Figure~1 shows a partial energy level diagram of Sb I, including the levels measured in this work. Both terms in the ($^3$P)6s configuration are plotted in Fig.~1, but only three of the six terms ($^4$P, $^4$D, $^4$F, $^2$P, $^2$D and $^2$F) in the ($^3$P)5d configuration.

\begin{figure*} [ht]
\resizebox{12cm}{!}{\includegraphics{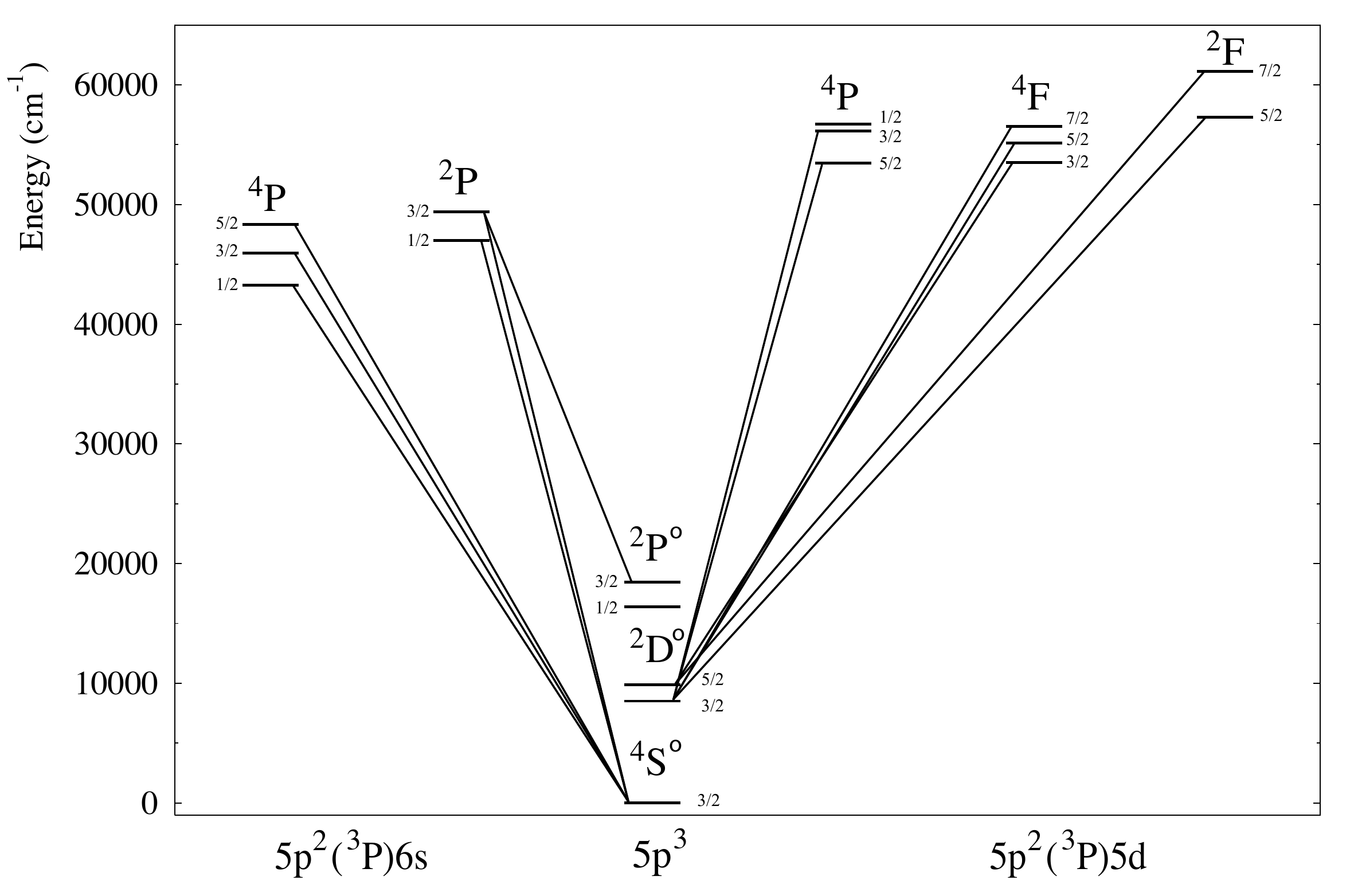}}
 \caption{Partial energy level diagram of Sb I including the levels measured in this work. The lines connecting the levels show the excitation routes used for the LIF experiment.}
\end{figure*} 

\begin{figure} [ht]
\resizebox{8cm}{!}{\includegraphics{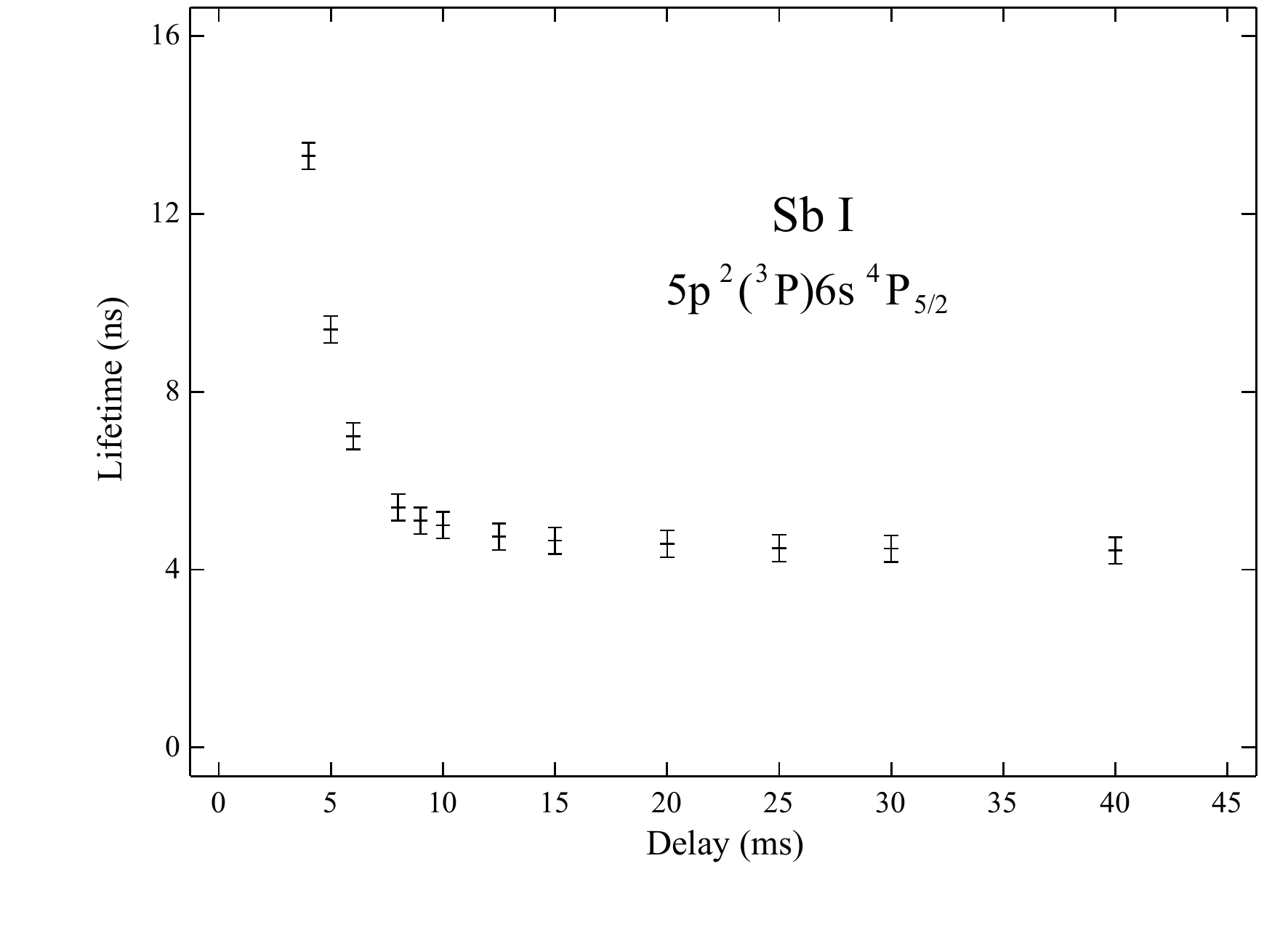}}
 \caption{Extracted lifetime of the 6s $^4$P$_{5/2}$ level as a function of the delay between the ablation and the excitation laser pulses, illustrating the severe saturation effect that occurs because the level is pumped from the ground state.}
\end{figure} 

\begin{figure} [ht]
\resizebox{8cm}{!}{\includegraphics{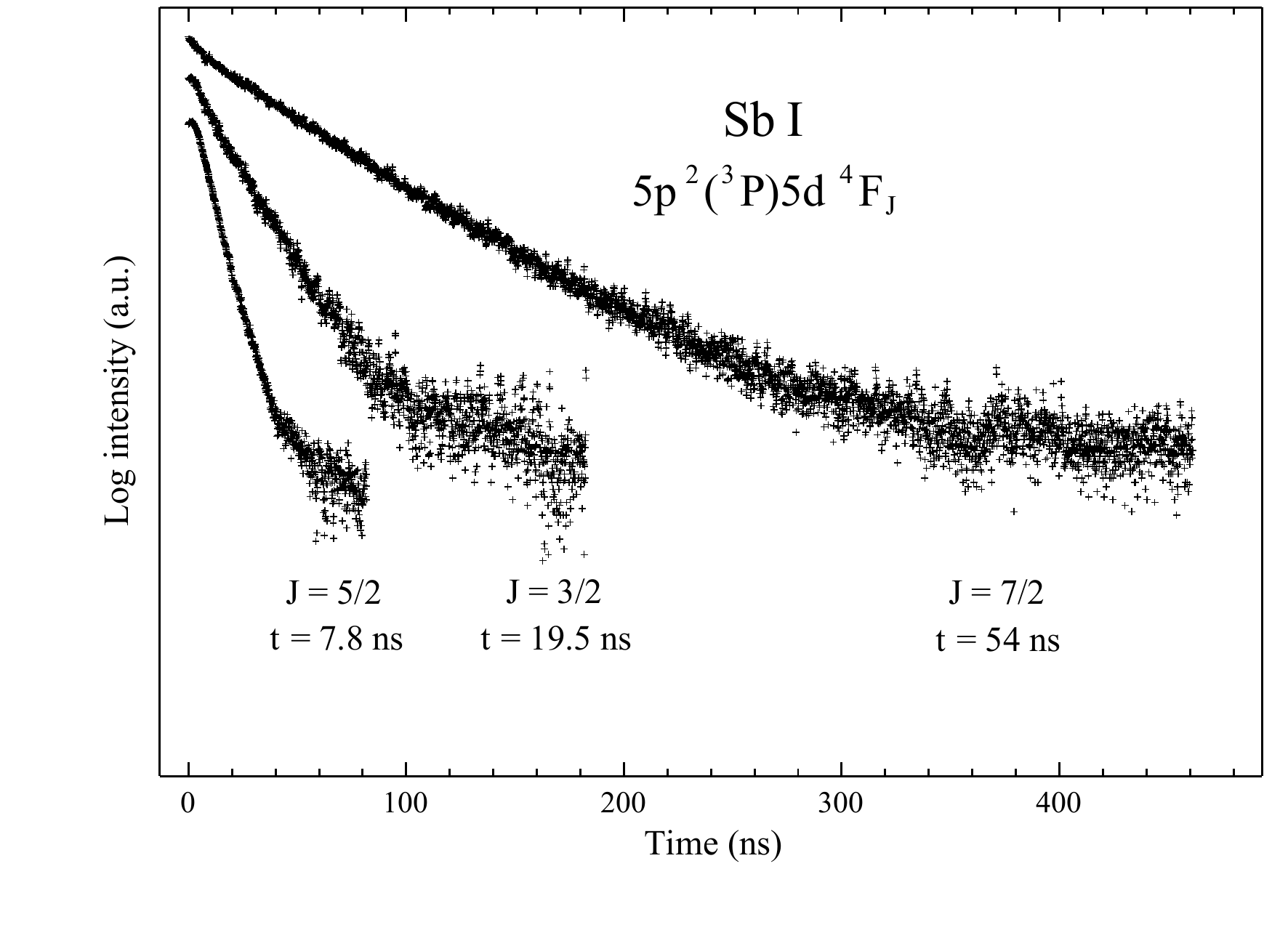}}
 \caption{Semi-logarithmic plot of the decay of the different fine structure levels in the 5d $^4$F term, illustrating the strong $J$-dependence found in the lifetimes. The signal registered before and during the excitation laser pulse is omitted, and the data is not background corrected. The decay of the metastable $^4$F$_{9/2}$ level could not be measured.}
\end{figure} 

\begin{figure}[ht]
\resizebox{7cm}{!}{\includegraphics{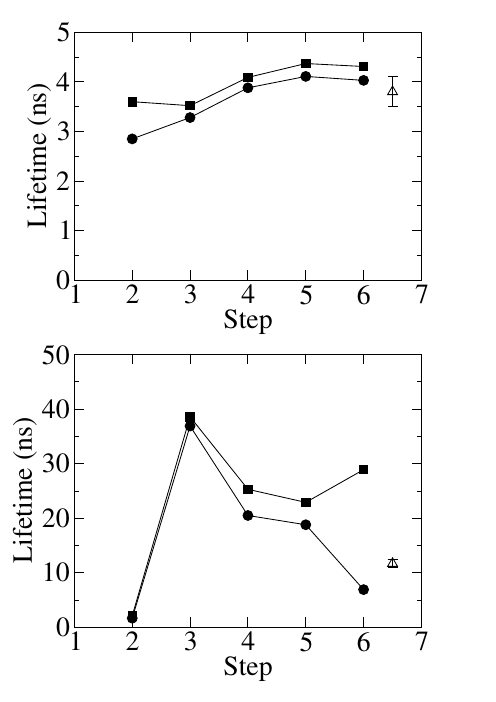}}
\caption{Examples of convergence monitoring in which the MCDHF lifetime (in ns) is
plotted as function of the computation step (see the text).
Filled circles and the filled squares represent the Babushkin and the
Coulomb gauge values, respectively. Upper panel: $\rm (^3P)5d~^2F_{7/2}$ level,
our TR-LIF measurement is shown with its error bar as an open triangle. Lower panel:
$\rm (^3P)5d~^4D_{1/2}$ level, the measurement of Osherovich \& Tezikov \cite{OT77} 
is shown with its error bar as an open triangle.}
\label{lifeconv}
\end{figure}

\begin{figure}[ht]
\resizebox{8cm}{!}{\includegraphics{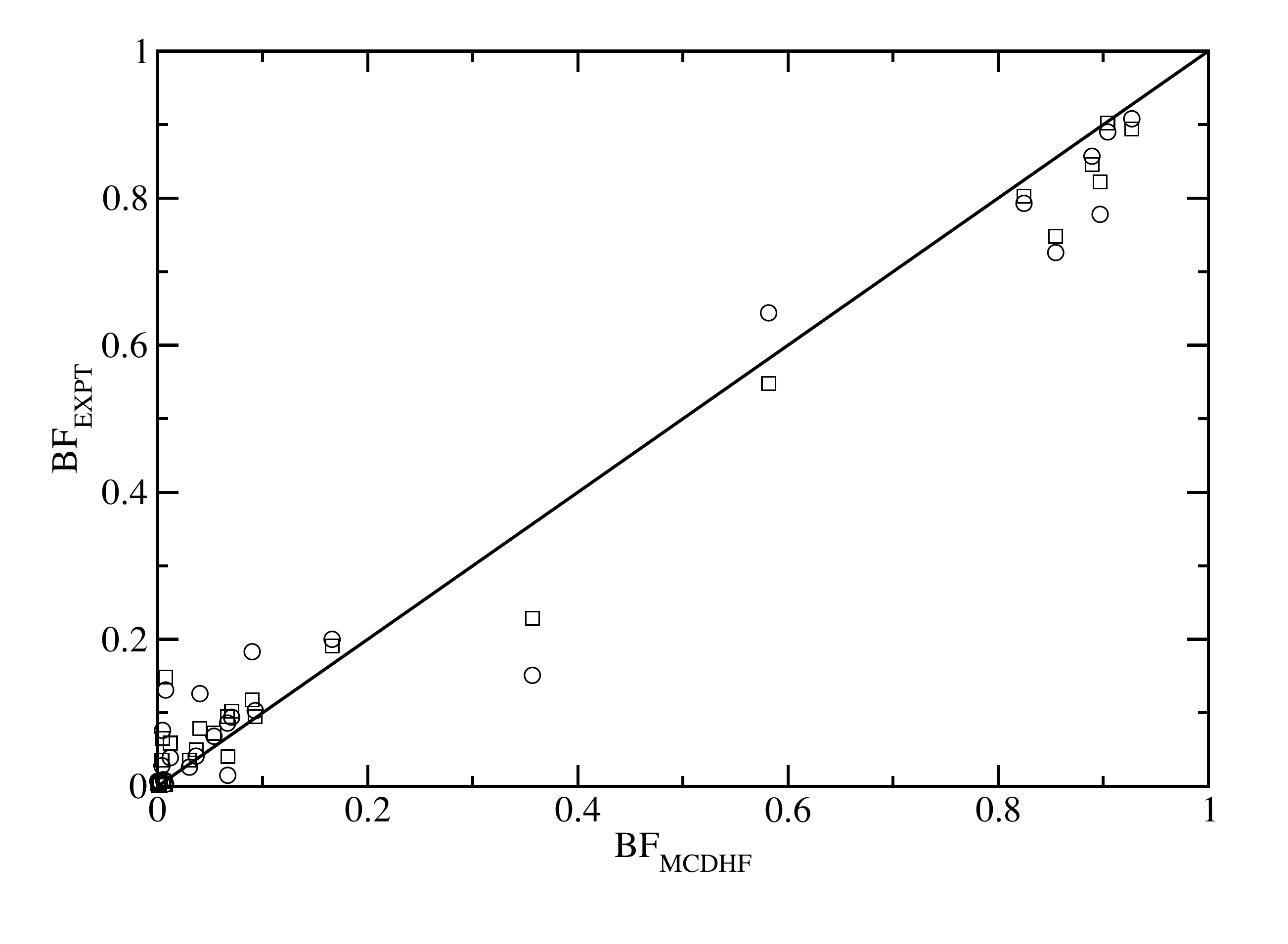}}
\caption{Comparison between corrected MCDHF (in the Babushkin gauge) and experimental 
branching fractions ($BF$). Circles:  Guern \& Lotrian \cite{GL}. Squares: Gonzalez {\it et al.} \cite{GOC}.
A straight line of equality is drawn.}
\label{BF}
\end{figure}

\begin{figure}[ht]
\resizebox{8cm}{!}{\includegraphics{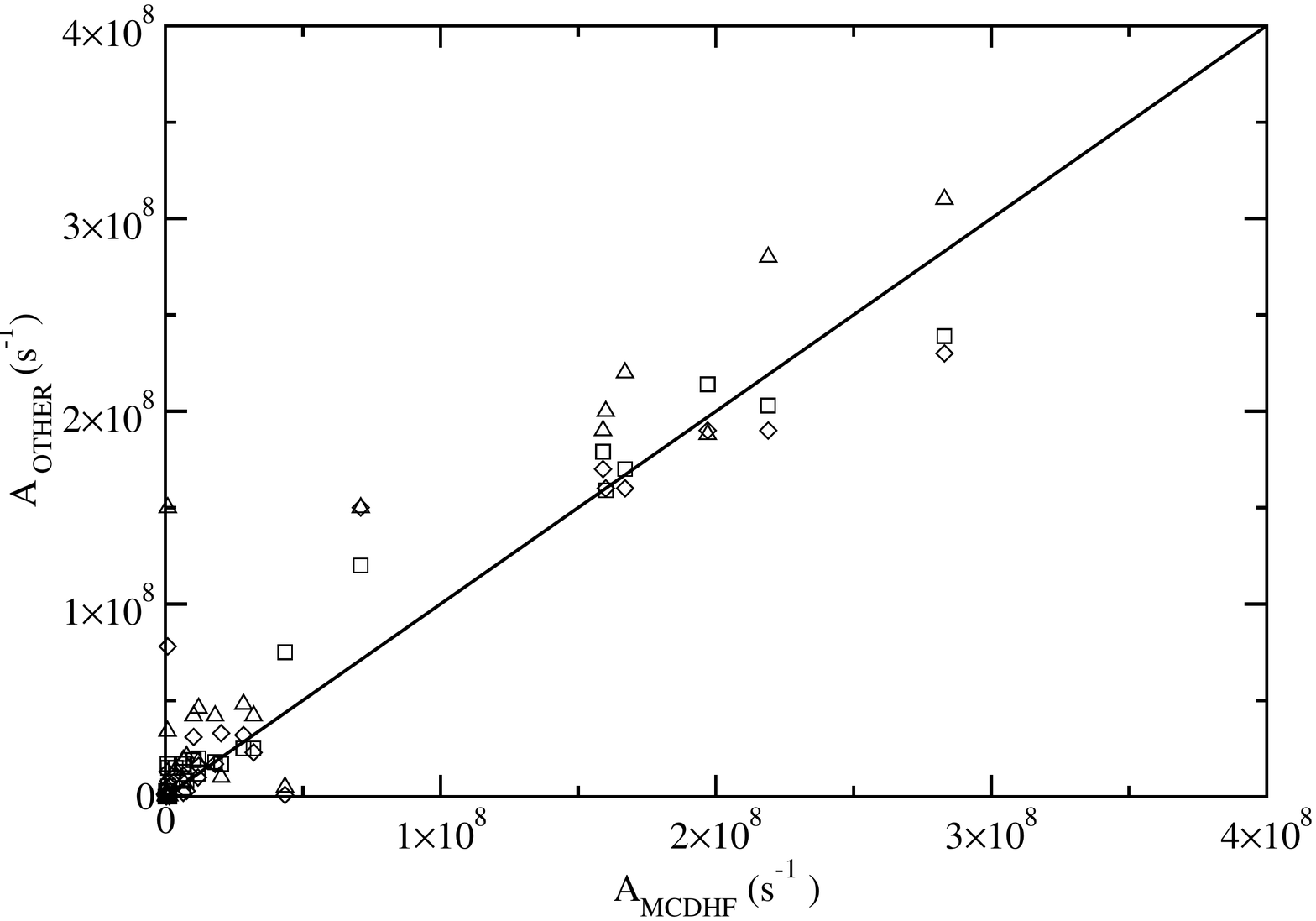}}
\caption{Comparison between corrected MCDHF transition probabilities (in the Babushkin gauge)
and theoretical values taken from the literature.
Squares: HF+MC calculation \cite{GOC}. Diamonds: 
MC+CI calculation (in the Babushkin gauge) \cite{BM}. 
Triangles: relativistic OHFS calculation (in the length gauge) \cite{HOL}.
A straight line of equality is drawn.}
\label{Aval}
\end{figure}

\section{Lifetime measurements}

The lifetimes were measured at the Lund High Power Laser Facility, Sweden, using the TR-LIF technique. For a detailed description of the instrumentation, see \cite{BFHLPW88}, \cite{XJZDSBQ03} and \cite{XPSBMPB04}. 

A laser-produced plasma was created by focusing a 10 Hz frequency doubled Nd:YAG laser onto a rotating antimony target, placed inside a vacuum chamber with a pressure of $\sim 10^{-5}$ millibar. The produced plasma contained free antimony atoms in the ground and excited states. Antimony atoms in low lying levels were selectively excited to the states under investigation and the subsequent fluorescence light was monitored as a function of time.  

The atoms were excited by crossing the ablation plasma 1 cm above the target with a (typically) 1.2 ns duration pulse from a dye laser pumped by a Nd:YAG laser. The pump laser pulses were shortened using stimulated Brillouin scattering. The desired wavelengths were obtained using a DCM dye, making the laser tunable between 600 and 660 nm. BBO and KDP doubling and tripling crystals were used to reach shorter wavelength. In addition, a hydrogen Raman shifter was used to change the frequency of the light by one or more Stokes or anti-Stokes shifts.  The delay between the ablation and excitation pulses was tuned to make the plasma contain only slow atoms, decreasing flight-in and flight-out effects and collisional quenching. For all 6s-levels, except $^2$P$_{3/2}$, the pumping laser was tuned to excite from the ground $^4$S$^{\circ}_{3/2}$ state (Fig. 1). Because of the strong reabsorption from the ground state, saturation effects turned out to be very important resulting in far too long experimental 6s lifetimes unless there was a long delay between the ablation and excitation laser pulses, leading to a plasma of sufficiently low-density. This effect is clearly illustrated in Fig. 2. In the case of 6s $^2$P$_{3/2}$ level, the excitation was from the 5p$^3$~$^2$P$^{\circ}_{3/2}$ level where no such effect could be seen and a typical delay of around 5 $\mu$s was used.

The fluorescence signal was focused on a 1/8 m monochromator, equipped with a PMT microchannel plate (Hamamatsu R1564) to detect the fluorescence signal. The laser pulse was detected separately with a photodiode. The fluorescence signal and the temporal shape of the laser pulse were recorded simultaneously with a transient digitizer and averaged over 1000 laser pulses. 

The short lifetimes were extracted by least square fitting of the fluorescence signal with a single exponential convoluted with the shape of the laser pulse, and a background function. Long lifetimes ($\tau > 10$\,ns) were extracted by fitting a single exponential and a background function to the fluorescence signal after the excitation pulse had died off completely. When possible we used different channels for excitation and detection of the fluorescence and more than one pump and/or detection channel as seen in Table~I. However, in some cases only one channel was accessible, and the pumping and detection had to be done in the same channel. 

In Table~I we report the excitation and detection (fluorescence) wavelengths, and the scheme used to excite the levels. The reported lifetimes in Table~II are averages of 10 to 20 lifetime curves and the uncertainties quoted include both statistical and systematic uncertainties. For comparison we give values from previous experiments found in the literature. The fourth column gives our theoretical lifetimes calculated with the MCDHF model and corrected by the experimental transition
energies of Hassini {\it et~al.} \cite{HARVW88}.     

\section{Calculations in Sb~I}

In order to calculate the lifetimes of the 6s and 5d levels measured in this work, we have used the MCDHF method implemented in the
GRASP2K computer package \cite{JFG}. In this method, the atomic state functions
(ASF), $\Psi(\gamma JM_J)$, are expanded in linear combinations of configuration state functions
(CSF), $\Phi(\alpha_i JM_J)$, according to
\begin{equation}
\Psi(\gamma JM_J)=~\sum_i c_i \Phi(\alpha_i JM_J).
\end{equation}

The CSFs are in turn linear combinations of Slater determinants constructed from monoelectronic
spin-orbitals of the form:
\begin{equation}
\varphi_{n\kappa m}(r,\theta,\phi)=~\frac{1}{r}
\begin{pmatrix} P_{n\kappa}(r) \chi_{\kappa m}(\theta,\phi) \\
 iQ_{n\kappa}(r) \chi_{-\kappa m}(\theta,\phi)
 \end{pmatrix}
\end{equation}
where $P_{n\kappa}(r)$ and $Q_{n\kappa}(r)$ are, respectively, the large and the small
component of the radial wave functions, and the angular functions $\chi_{\kappa m}(\theta,\phi)$
are the spinor spherical harmonics \cite{GRA}. The $\alpha_i$ represent all the
one-electron and intermediate quantum numbers needed to completely define the CSF.
$\gamma$ is usually chosen as the $\alpha_i$ corresponding to the CSF with the largest
weight $\arrowvert c_i \arrowvert ^2$. The quantum number $\kappa$ is given by
\begin{equation}
\kappa =~\pm (j+\frac{1}{2})
\label{kappa}
\end{equation}
where $j$ is the electron total angular momentum. 
 The sign before the parenthesis in Eq.~(\ref{kappa}) corresponds
to the coupling relation between the electron orbital momentum, $\ell$, and its spin, i.e.
\begin{equation}
\ell =~j \pm \frac{1}{2}.
\end{equation}
The radial functions $P_{n\kappa}(r)$ and 
$Q_{n\kappa}(r)$ are numerically represented on a logarithmic grid and are required to be
orthonormal within each $\kappa$ symmetry. In the MCDHF variational procedure, the radial
functions and the expansion coefficients $c_i$ are optimized to self-consistency.

We considered the active space (AS) method for building the MCDHF multiconfiguration
expansions. The latter are produced by exciting the electrons from the reference configurations
to a given set of orbitals. The rules adopted for generating the configuration space differ according
to the correlation model being used. Within a given correlation model, the AS of orbitals spanning
the configuration space is increased to monitor the convergence of the total energies and the
transition probabilities.

The MCDHF calculations have been carried out in 6 steps. 

In the first step, the core orbitals,
i.e. 1s to 4d, together with the 5s and 5p orbitals, have been optimized. 
The 6 CSF belonging to the ground configuration 5s$^2$5p$^3$ were retained in the configuration space. The energy functional was built within the framework 
of the Average Level (AL) option \cite{GRA}. 

The second step consisted in increasing the
configuration space by considering all the 62 CSF belonging to the following configurations:
5s$^2$5p$^3$+5s$^2$5p$^2$\{6s,6p,5d\}$^1$. The 6s, 6p and 5d orbitals
have been optimized keeping the others fixed to their values of the first step. The AL option was
chosen to build the energy functional. 

In the third step, the configuration space has been
extended to 13640 CSF by considering the single and double virtual electron excitations to the
active orbital set \{5s,5p,5d,6s,6p,6d\} from the multireference configurations
5s$^2$5p$^3$+5s$^2$5p$^2$\{6s,6p,5d\}$^1$. Only the 6d orbital has been optimized fixing
all the others to the values of the preceeding step using an energy functional built from the
lowest 62 ASF within the framework of the Extended Optimal Level (EOL) option \cite{GRA}. 
One can notice that, from this step of the computation and on, core-valence and core-core
correlations are also considered through the single and double excitations of the 5s core electrons,
respectively.

The last three steps consisted in extending further the configuration
space by adding to the active set of the preceeding steps the following orbitals : 
7s, 7p and 7d in the fourth step giving rise to 41 603 CSF; 8s, 8p and 8d in the fifth step generating
85 130 CSF; and finally, 4f in the last step with 118 912 CSF generated. In these steps, only
the added orbitals have been optimized, the others being fixed using the same energy functional as in
the third step; also, single and double virtual electron excitations from the same multireference
configurations as in the third step have been used to generate the configuration spaces.
Further orbital additions to the active set as well as further opening of the core to include more
core-valence and core-core correlations have been prevented by the memory limitations
of our computer.

Two examples of convergence monitoring are shown in Fig.~\ref{lifeconv}. The MCDHF lifetimes
of the $\rm (^3P)5d~^2F_{7/2}$ and $\rm (^3P)5d~^4D_{1/2}$ levels as computed in the Babushkin 
and  the Coulomb gauges are plotted as function of the computation step. The values of the
last step can be compared to our TR-LIF measurement for the $\rm (^3P)5d~^2F_{7/2}$
level and to the delayed coincidence measurement of Osherovich \& Tezikov \cite{OT77}
for the $\rm (^3P)5d~^4D_{1/2}$ level. As one can see, in the case of $\rm (^3P)5d~^2F_{7/2}$,
the calculated lifetime converges in both gauges to the experimental value whereas 
it clearly diverges in the case of $\rm (^3P)5d~^4D_{1/2}$. The divergence of the 
MCDHF lifetime was also noticed for the  $\rm (^3P)5d~^4D_{5/2}, ^4P_{1/2}$ levels
not shown here. These three 5d states are probably not properly represented due to missing correlations, e.g.
with $n$f and $n$g ($n \ge 5$) orbitals.

In Table~\ref{life}, we report the MCDHF energies of the levels for which
lifetime measurements are available. The corresponding MCDHF lifetimes and transition probabilities of main decay branches (presented in Table~\ref{trans}) 
have been corrected by the experimental transition energies of Hassini {\it et~al.} \cite{HARVW88} as already stated in the previous section. Values affected by convergence problems are marked with
an asterisk.

All the computed values reported here are without Breit and QED
interactions. Rough estimates of the importance of these interactions on
the lifetimes have been established
in the second step of the computation; the lifetimes changed by less than
5\%.

\section{Discussion}

The 12 measured levels in Sb~I are reported in Table~II. Among them,  7 levels had not been previously considered. 

Comparing the TR-LIF lifetimes from this work with values in the literature, we agree within the uncertainties with Belin {\it et~al.} \cite{BGHR74} and Andersen  {\it et~al.} \cite{AWS74}. However, there is a clear discrepancy when comparing to Osherovich \& Tezikov \cite{OT77}. 

It is interesting to note that the lifetimes within the 5d $^4$F term differ by more than a factor of six, see Figure 3. Only the three levels with the lowest $J$-values (3/2, 5/2 and 7/2) are measured in this term. The level with the highest J-value, $^4$F$_{9/2}$, is metastable since there are no odd levels with $J \geq 7/2$ below 57 000 cm$^{-1}$ to which it can decay through an electric dipole transition. The shortest lifetime is that of $J=5/2$ (7.8 ns). This can be explained by the large mixing with 5d $^4$P$_{5/2}$ which opens up a strong decay channel to the ground level $^4$S$^{\circ}_{3/2}$, decreasing the lifetime. The level 5d $^4$F$_{3/2}$ mixes with 5d $^4$P$_{3/2}$, but not as much as in the previous case, and the transition to $^4$S$^{\circ}_{3/2}$ is weaker giving the longer lifetime of 19.5 ns. The level 5d $^4$F$_{7/2}$ has a too large $J$-value to mix with the 5d $^4$P since level mixing can only occur between levels with the same $J$-value. However, this level is mixed with $^2$F, opening up the channel down to 5p$^3$ $^2$D$^{\circ}_{5/2}$. The experimental lifetimes for the levels in the $^4$F term are nicely reproduced by the theoretical calculation.

The MCDHF level energies are compared with the experimental data of Hassini {\it et~al.} \cite{HARVW88} in Table II (see column 3). The agreement is excellent.

A comparison between experimental and calculated lifetimes is also shown in Table~\ref{life}. The calculated level energies appear systematically but consistently lower than the experimental results by a few hundreds cm$^{-1}$. 
%
%
Our MCDHF calculations are in agreement with our LIF measurements and with the values of Belin {\it et~al.} \cite{BGHR74} and Andersen  {\it et~al.} \cite{AWS74}, whereas the values of Osherovich \& Tezikov \cite{OT77} seem to have larger uncertainties than quoted.

A good agreement is found between our MCDHF and
TR-LIF lifetimes with most of the calculated values being within the error bars. Moreover,
one can observe the good agreement between the Coulomb and Babushkin gauges for most of the MCDHF lifetimes.
This is a necessary but not a sufficient condition for the accuracy of the calculations. It should be emphasized that
this agreement was not observed in the earlier MC+CI \cite{BM} and OHFS calculations 
\cite{HOL}. Discrepancies between the two gauges reaching a factor of two and more are however observed for the three high excitation levels at 56 698, 57 597 and 58 862 cm$^{-1}$ due to the convergence problems mentioned in the previous
section. Albeit the HF+MC calculations of Gonzalez {\it et al.} \cite{GOC} show an excellent
agreement with our measurements for the 6s levels, we have a completely different situation 
as far as the 5d levels are concerned. The latters were not published in \cite{GOC}
and we have therefore calculated them using the model kindly provided by these authors.

Regarding the previous measurements, the beam-foil data \cite{AWS74}
 and the Hanle measurements \cite{BGHR74}
have been confirmed by our TR-LIF lifetimes, whilst the beam-foil values for the 
6s~$^2$D$_{3/2,5/2}$ levels are significantly shorter than our MCDHF predictions.
The problem is probably originating from the MCDHF values because these two 6s $^2$D  levels are strongly mixed with 5d levels (which is not the case for the other levels of the same configuration), the dominant percentages being less than 25\% in jj coupling for these two levels. The better agreement theory-experiment observed for the HF+MC is probably due to the fact that the authors of Ref. \cite{GOC} used a semi-empirical optimization procedure, the results of the present work being purely theoretical.

The delay coincidence measurements of Osherovich \& Tezikov \cite{OT77} and of Tezikov \cite{T78}
are, on the other hand, found to be significantly shorter or longer than our experimental
and theoretical lifetimes. This is also true for the 6s~$^2$P$_{1/2}$ level where the only previous experimental lifetimes are obtained by the delayed coincidence technique \cite{OT77,T78}. However, we note that in their branching fraction measurements  Guern \& Lotrian \cite{GL}  proposed a lifetime of 3.7$\pm$0.7 ns for
this level from a Boltzmann plot. This value is in close agreement with both our TR-LIF and MCDHF results.

The corrected MCDHF transition probabilities, along with the corresponding oscillator strengths and branching fractions, are reported in Table~\ref{trans} for the main decay transitions depopulating
the even levels presented in Table~\ref{life}. 
A good agreement between the gauges is generally found for the strong lines except for the lines depopulating the three high excitation levels ($^3$P)5d $^4$D$_{1/2,5/2}$ and ($^3$P)5d $^4$P$_{1/2}$ due to the convergence problems mentioned in the
previous section. 

In Fig.~\ref{BF}, our calculated branching fractions
(in the Babushkin gauge) as reported in Table~\ref{trans} are compared to the measurements 
of Guern \& Lotrian \cite{GL} and of Gonzalez {\it et~al.} \cite{GOC}. The averages of the ratios $BF_{EXPT}/BF_{MCDHF}$ for the strong branches ($BF_{MCDHF} \ge 0.1$) excluding the values for the decay transition 6s~$^2$D$_{3/2}$$-$5p$^3$~$^4$S$^o_{3/2}$, for which there is a disagreement of up to
a factor of two, are respectively 0.99$\pm$0.12 for Guern \& Lotrian \cite{GL} and 
0.97$\pm$0.08 for Gonzalez {\it et~al.} \cite{GOC}
(the quoted uncertainties being the standard deviation to the mean) showing 
a reasonable agreement between the three data sets. 
Concerning the $BF$ of the transition 6s~$^2$D$_{3/2}$$-$5p$^3$~$^4$S$^o_{3/2}$, 
the strong disagreement between theory and experiment is due to the inaccuracy of the MCDHF
$A$-value for the decay branch 6s~$^2$D$_{3/2}$$-$5p$^3$~$^2$P$^o_{3/2}$
($9.33\times10^5$ s$^{-1}$ in the Babushkin gauge and $1.44\times10^7$ s$^{-1}$ in the
Coulomb gauge) that is involved in the calculation of the $BF$ of the decay branch
6s~$^2$D$_{3/2}$$-$5p$^3$~$^4$S$^o_{3/2}$.

Our corrected MCDHF transition probabilities (in the Babushkin gauge) are compared with the available theoretical data in the literature 
(\cite{BM},\cite{HOL},\cite{GOC}) in Fig.~\ref{Aval}.
The relativistic OHFS $A$-values of Holmgren \cite{HOL} for  the strong lines ($A \ge 10^8$ s$^{-1}$)
are definitely too high.  The agreement is somewhat better with the other two theoretical methods \cite{GOC} \cite{BM}
for the strong lines if we except the transition 
6s~$^2$P$_{1/2}$$-$5p$^3$~$^2$D$^o_{3/2}$ with $A_{MCDHF}= 2.83\times10^8$
(Babushkin)/$2.72\times10^8$(Coulomb) s$^{-1}$. 
The HF+MC \cite{GOC} and MC+CI  \cite{BM} results are predicted too low (length-gauge) for that transition, i.e.
$2.39\times10^8$ and $2.30\times10^8$ s$^{-1}$, respectively. Unfortunately, both authors
did not publish their velocity-gauge $A$-values. Nonetheless, all these three "old" calculations
are expected to be less accurate than the present results due to the consideration of configuration interaction effects in a more limited way. 

\begin{acknowledgments}
This work was financially supported by the Integrated Initiative of Infrastructure Project LASERLAB-EUROPE, contract  RII3-CT-2003-506350, the Swedish Research Council through the Linnaeus grant, the Knut and Alice Wallenberg Foundation and the Belgian FRS-FNRS. EB, PQ and PP are, respectively, Research Director, Senior Research Associate and Research Associate of the FRS-FNRS. We thank our Spanish colleagues (A.M. Gonzalez, M. Ortiz, and J. Campos) for providing us with some details about their calculations.
\end{acknowledgments}

\begin{table*}
\caption{\label{tab:table1}Pumping scheme for Sb I levels. The first three columns are the pumped level under investigation, column 4 the lower level, column 5 the pumping transition, column 6 the fluorescence channel, column 7 the level to which the fluorescence decays. The last column describes the technique used to produce desired frequency of the pumping radiation.}
\begin{ruledtabular}
\begin{tabular}{lllrcccc}
Config.$^a$ & Level$^a$ & $E$$^a$ / cm$^{-1}$& Pump level & Pump $\lambda$ / \AA & Fluorescence $\lambda$ / \AA & Lower level & Method \\

\hline

5p$^2$($^3$P)6s  &	$^4$P$_{1/2}$ & 43249.337 & $^4$S$^{\circ}_{3/2}$		& 2311.463 	& 2878			& $^2$D$^{\circ}_{3/2}$ 				& 3$\omega+$S\\
5p$^2$($^3$P)6s	&	$^4$P$_{3/2}$ & 45945.340 & $^4$S$^{\circ}_{3/2}$   	& 2175.818  	& 2176			& $^4$S$^{\circ}_{3/2}$				& 3$\omega$ \\
5p$^2$($^3$P)6s	&	$^4$P$_{5/2}$ & 48332.424 & $^4$S$^{\circ}_{3/2}$   	& 2068.344 	& 2068 			& $^4$S$^{\circ}_{3/2}$				& 3$\omega$ \\
&&&&&&&\\
5p$^2$($^3$P)6s	&	$^2$P$_{1/2}$ & 46991.058 & $^4$S$^{\circ}_{3/2}$   	& 2127.39 	& 2599			& $^2$D$^{\circ}_{3/2}$ 				& 3$\omega$ \\
5p$^2$($^3$P)6s	&	$^2$P$_{3/2}$ & 49391.133 & $^4$S$^{\circ}_{3/2}$ 		& 2024.00 	& 2529/3232		& $^2$D$^{\circ}_{5/2}$/$^2$P$^{\circ}_{3/2}$ 	& 3$\omega+$2S \\ 
                &               &           & $^2$P$_{3/2}$     & 3232.53   & 2529/3232		& 
                $^2$D$^{\circ}_{5/2}$/$^2$P$^{\circ}_{3/2}$ 	& 2$\omega$ \\ 
 &&&&&&&\\                
5p$^2$($^3$P)5d	&	$^4$P$_{5/2}$ & 53442.967 & $^2$D$^{\circ}_{3/2}$ 	& 2225.00		& 2290			& $^2$D$^{\circ}_{5/2}$	 			& 3$\omega$ \\
5p$^2$($^3$P)5d	&	$^4$P$_{3/2}$ & 56151.802 & $^2$D$^{\circ}_{3/2}$ 	& 2098.00		& 2098			& $^2$D$^{\circ}_{3/2}$ 				& 3$\omega$\\
 &&&&&&&\\
 5p$^2$($^3$P)5d	&	$^4$F$_{3/2}$ & 53527.956 & $^2$D$^{\circ}_{3/2}$  	& 2220.75		& 2850/2690/2280	& $^2$P$^{\circ}_{1/2}$/$^2$P$^{\circ}_{3/2}$/$^2$D$^{\circ}_{3/2}$							& 3$\omega$ \\
5p$^2$($^3$P)5d	&	$^4$F$_{5/2}$ & 55120.943 & $^2$D$^{\circ}_{3/2}$  	& 2144.86		& 2208			& $^2$D$^{\circ}_{5/2}$ 				& 3$\omega$ \\
5p$^2$($^3$P)5d	&	$^4$F$_{7/2}$ & 56528.132 & $^2$D$^{\circ}_{5/2}$  	& 2141.83		& 2142			& $^2$D$^{\circ}_{5/2}$				& 3$\omega$\\
 &&&&&&&\\
 5p$^2$($^3$P)5d	&	$^2$F$_{5/2}$ & 57287.052 & $^2$D$^{\circ}_{3/2}$  	& 2049.57		& 2050			& $^2$D$^{\circ}_{3/2}$ 				& 3$\omega$\\
5p$^2$($^3$P)5d	&	$^2$F$_{7/2}$ & 61125.741 & $^2$D$^{\circ}_{5/2}$ 		& 1950.39		& 1950			& $^2$D$^{\circ}_{5/2}$				& 3$\omega+$AS\\

\end{tabular}
\end{ruledtabular}
\begin{list}{}{}
\item[$^{\mathrm{a}}$] From Hassini et~al.~\cite{HARVW88}.\\
2$\omega$ and 3$ \omega$ mean the second and the third harmonic, S and AS are written for the Stokes and anti-Stokes components of the Raman scattering.
\end{list}
\end{table*}

\begin{table*}
\caption{\label{life}Comparison between experimental and calculated radiative lifetimes ($\tau$ in ns) in Sb~{\sc i}. Values marked with an asterisk were
affected by convergence problems (see the text).}
\begin{ruledtabular}
\begin{tabular}{lrrrrrr}
&&&&& \multicolumn{2}{c}{Other studies}\\
\cline{6-7}
\multicolumn{1}{c}{Level$^a$}&\multicolumn{1}{c}{E$_{exp}$(cm$^{-1}$)$^a$}&
\multicolumn{1}{c}{E$_{MCDHF}$(cm$^{-1}$)$^b$}&
\multicolumn{1}{c}{$\tau_{MCDHF}$ (ns)$^b$} &
\multicolumn{1}{c}{$\tau_{LIF}$ (ns)$^c$} &
\multicolumn{1}{c}{$\tau_{exp}$ (ns)}
&\multicolumn{1}{c}{$\tau_{calc}$ (ns)}\\
\hline
($^3$P)6s $^4$P$_{1/2}$&43249.337&42917&5.2(5.4)&5.3$\pm$0.2&5.0$\pm$0.4$^d$&5.4$^g$\\
&&&&&4.3$\pm$0.4$^f$&5.4(8.3)$^h$\\
&&&&&&4.1(7.5)$^i$\\
($^3$P)6s $^4$P$_{3/2}$&45945.340&45419&5.6(5.4)&5.3$\pm$0.2&4.9$\pm$0.5$^d$&5.1$^g$\\
&&&&&5.1$\pm$0.6$^e$&5.3(7.8)$^h$\\
&&&&&4.6$\pm$0.5$^f$&4.5(8.7)$^i$\\
($^3$P)6s $^4$P$_{5/2}$&48332.424&47797&4.6(4.9)&4.6$\pm$0.3&4.8$\pm$0.4$^d$&4.3$^g$\\
&&&&&4.8$\pm$0.6$^e$&5.0(6.9)$^h$\\
&&&&&&4.5(8.9)$^i$\\
&&&&&&\\
($^3$P)6s $^2$P$_{1/2}$&46991.058&46783&3.2(3.2)&3.7$\pm$0.2&5.2$\pm$0.5$^f$&3.7$^g$\\
&&&&&5.5$\pm$0.7$^f$&2.7(3.7)$^i$\\
($^3$P)6s $^2$P$_{3/2}$&49391.133&49217&3.9(3.6)&3.8$\pm$0.2&4.0$\pm$0.3$^d$&3.9$^g$\\
&&&&&4.3$\pm$0.4$^f$&4.0(5.0)$^h$\\
&&&&&&2.6(3.6)$^i$\\
&&&&&&\\
($^1$D)6s $^2$D$_{3/2}$&55232.963&55071&8.2(6.1)&&3.7$\pm$0.4$^d$&4.4$^g$\\
&&&&&&4.3(6.3)$^h$\\
&&&&&&3.0(5.0)$^i$\\
($^1$D)6s $^2$D$_{5/2}$&55728.268&55435&5.5(5.9)&&3.8$\pm$0.3$^d$&4.7$^g$\\
&&&&&3.9$\pm$0.3$^j$&5.2(7.1)$^h$\\
&&&&&&3.2(5.7)$^i$\\
&&&&&&\\
($^3$P)5d $^4$P$_{5/2}$&53442.967&53059&6.1(8.9)&7.0$\pm$1.0&&3.8$^g$\\
($^3$P)5d $^4$P$_{3/2}$&56151.802&55683&5.3(5.5)&6.0$\pm$0.4&4.1$\pm$0.2$^j$&3.7$^g$\\
($^3$P)5d $^4$P$_{1/2}$&56698.608&56196&4.9(10.4)*&&10.3$\pm$0.7$^f$&5.5$^g$\\
&&&&&&\\
($^3$P)5d $^4$F$_{3/2}$&53527.956&53438&23.4(23.2)&19.5$\pm$1.5&&11.3$^g$\\
($^3$P)5d $^4$F$_{5/2}$&55120.943&54928&9.6(10.6)&7.8$\pm$0.4&&12.6$^g$\\
($^3$P)5d $^4$F$_{7/2}$&56528.132&56265&60.8(59.8)&54$\pm$3&&30.4$^g$\\
&&&&&&\\
($^3$P)5d $^2$P$_{3/2}$&56733.162&56542&8.5(9.8)&&3.5$\pm$0.4$^j$&3.6$^g$\\
&&&&&&\\
($^3$P)5d $^2$F$_{5/2}$&57287.052&57235&9.9(9.0)&9.5$\pm$0.5&&4.4$^g$\\
($^3$P)5d $^2$F$_{7/2}$&61125.734&60941&3.8(4.1)&3.8$\pm$0.3&&2.7$^g$\\
&&&&&&\\
($^3$P)5d $^4$D$_{1/2}$&57597.203&57403&6.5(27.7)*&&11.7$\pm$0.8$^f$&9.4$^g$\\
($^3$P)5d $^4$D$_{5/2}$&58862.889&58684&6.3(13.7)*&&10.4$\pm$0.4$^f$&10.6$^g$\\
&&&&&&\\

\end{tabular}
\end{ruledtabular}
\begin{small}
\begin{itemize}
\item[$^a$] Hassini {\it et al.} \cite{HARVW88}.
\item[$^b$] MCDHF calculation (this work).
The lifetimes are corrected from the experimental transition energies.
a(b) stands for Babushkin(Coulomb).
\item[$^c$] TR-LIF measurements (this work).
\item[$^d$] Beam-foil spectroscopy \cite{AWS74}.
\item[$^e$] Hanle method \cite{BGHR74}.
\item[$^f$] Delayed coincidence technique \cite{OT77}.
\item[$^g$] HF+MC calculation \cite{GOC}. The values for the 5d levels have been
calculated by us using the Hartree-Fock model kindly provided by the authors.
\item[$^h$] MC+CI calculation \cite{BM}. a(b) stands for Babushkin(Coulomb).
\item[$^i$] Relativistic OHFS calculation \cite{HOL}. a/b stands for length(velocity).
\item[$^j$] Delayed coincidence technique \cite{T78}.
\end{itemize}
\end{small}
\end{table*}

\begin{table*}
\caption{\label{trans}Transition probabilities ($A_{ki}$), weighted oscillator strengths
($gf_{ik}$) and branching fractions ($BF$) of decay transitions depopulating the even levels
 reported in Table~\ref{life}. Only transitions with wavelengths $\lambda$ $\leq$ 2500 nm and A$_{ki}$ $\geq$ 1.00E+06 are quoted. r is the ratio between Babushkin and Coulomb gauges.}
\begin{ruledtabular}
\begin{tabular}{llrcccc}
\multicolumn{1}{c}{Upper Level$^a$}&\multicolumn{1}{c}{Lower Level$^a$}&
\multicolumn{1}{c}{$\lambda$(nm)$^a$}&
\multicolumn{1}{c}{$A_{ki}$ (s$^{-1}$)$^b$} &
\multicolumn{1}{c}{r}&
\multicolumn{1}{c}{$gf_{ik}^b$} &
\multicolumn{1}{c}{$BF^b$}\\
\hline
6s $^4$P$_{1/2}$	&	5p$^3$ $^4$S$^o_{3/2}$	&	231.146	&	1.60E+08		&1.045&	2.56E-01		&	8.25E-01		\\
	&	5p$^3$ $^2$D$^o_{3/2}$	&	287.791	&	3.21E+07	 	&1.092&	7.99E-02		&	1.66E-01		\\
	&	5p$^3$ $^2$P$^o_{3/2}$	&	403.354	&	1.47E+06	 	&1.610&	7.17E-03	 	 	&	7.59E-03	 	\\
	&		&		&				&				&				\\
6s $^4$P$_{3/2}$	&	5p$^3$ $^4$S$^o_{3/2}$	&	217.582	&	1.59E+08	&0.946	&	4.51E-01	 	&	8.89E-01	 	\\
	&	5p$^3$ $^2$D$^o_{3/2}$	&	267.063	&	6.55E+06	 	&1.224&	2.80E-02	 	&	3.67E-02	 	\\
	&	5p$^3$ $^2$D$^o_{5/2}$	&	276.993	&	1.19E+07	 	&0.930&	5.47E-02	 	&	6.66E-02	 	\\
	&		&		&				&				&				\\
6s $^2$P$_{1/2}$	&	5p$^3$ $^4$S$^o_{3/2}$	&	212.739	&	3.82E+06		&0.479&	5.18E-03	 	&	1.21E-02	 	\\
	&	5p$^3$ $^2$D$^o_{3/2}$	&	259.805	&	2.83E+08	 	&1.040&	5.72E-01	 	&	8.97E-01	 	\\
	&	5p$^3$ $^2$P$^o_{1/2}$	&	326.749	&	2.84E+07	 	&0.976&	9.08E-02	 	&	9.00E-02	 	\\
	&		&		&				&				&				\\
6s $^4$P$_{5/2}$	&	5p$^3$ $^4$S$^o_{3/2}$	&	206.834	&	1.97E+08	&1.070	&	7.60E-01	 	&	9.04E-01	 	\\
	&	5p$^3$ $^2$D$^o_{5/2}$	&	259.808	&	2.03E+07	 	&	1.009&1.23E-01 	&	9.29E-02	 	\\
	&		&		&				&				&				\\
6s $^2$P$_{3/2}$	&	5p$^3$ $^4$S$^o_{3/2}$	&	202.400	&	1.06E+06	 &0.270	&	2.60E-03	 	&	4.13E-03	 	\\
	&	5p$^3$ $^2$D$^o_{3/2}$	&	244.550	&	7.75E+06	 	&1.305&	2.78E-02	 	&	3.03E-02	 	\\
	&	5p$^3$ $^2$D$^o_{5/2}$	&	252.851	&	2.19E+08	 	&0.940&	8.40E-01 	&	8.55E-01	 	\\
	&	5p$^3$ $^2$P$^o_{1/2}$	&	302.981	&	1.81E+07	 	&1.000&	9.95E-02	 	&	7.05E-02	 	\\
	&	5p$^3$ $^2$P$^o_{3/2}$	&	323.249	&	1.03E+07	 	&0.520&	6.47E-02	 	&	4.03E-02	 	\\
	&		&		&				&				&				\\
5d $^4$P$_{5/2}$	&	5p$^3$ $^4$S$^o_{3/2}$	&	187.115	&	1.50E+08	 &1.471	&	4.72E-01	 	&	9.16E-01	 	\\
	&	5p$^3$ $^2$D$^o_{3/2}$	&	222.495	&	9.64E+06&1.148	 	&	4.30E-02	 	&	5.90E-02	 	\\
	&	5p$^3$ $^2$D$^o_{5/2}$	&	229.345	&	2.33E+06	 	&1.069&	1.10E-02	 	&	1.42E-02	 	\\
	&		&		&				&				&				\\
5d $^4$F$_{3/2}$	&	5p$^3$ $^4$S$^o_{3/2}$	&	186.818	&	9.04E+06	&0.615 	&	1.89E-02	 	&	2.11E-01	 	\\
	&	5p$^3$ $^2$D$^o_{3/2}$	&	222.075	&	8.72E+06	 	&0.765&	2.58E-02	 	&	2.04E-01	 	\\
	&	5p$^3$ $^2$D$^o_{5/2}$	&	228.899	&	7.31E+06	 	&1.417&	2.30E-02	 	&	1.71E-01	 	\\
	&	5p$^3$ $^2$P$^o_{1/2}$	&	269.225	&	4.46E+06	 	&1.143&	1.94E-02	 	&	1.04E-01	 	\\
	&	5p$^3$ $^2$P$^o_{3/2}$	&	285.111	&	1.30E+07	 &1.712&	6.32E-02	 	&	3.03E-01	 	\\
		&		&		&				&				&				\\
	5d $^4$F$_{5/2}$	&	5p$^3$ $^4$S$^o_{3/2}$	&	181.419	&	4.09E+07	 &1.786	&	1.21E-01	 	&	3.92E-01	 	\\
	&	5p$^3$ $^2$D$^o_{3/2}$	&	214.484	&	3.02E+07	 	&0.647&	1.25E-01	 	&	2.90E-01	 	\\
	&	5p$^3$ $^2$P$^o_{3/2}$	&	272.720	&	1.37E+07	 	&0.980&	9.15E-02	 	&	1.31E-01	 	\\
	&		&		&				&				&				\\
6s $^2$D$_{3/2}$	&	5p$^3$ $^4$S$^o_{3/2}$	&	181.051	&	4.35E+07	&0.737 	&	8.56E-02 	&	3.57E-01	 	\\
	&	5p$^3$ $^2$D$^o_{3/2}$	&	213.970	&	7.10E+07	 	&0.861&	1.95E-01	 	&	5.82E-01	 	\\
	&	5p$^3$ $^2$P$^o_{1/2}$	&	257.405	&	6.55E+06	 &1.065	&	2.60E-02	 	&	5.36E-02	 	\\
	&		&		&				&				&				\\
6s $^2$D$_{5/2}$	&	5p$^3$ $^2$D$^o_{3/2}$	&	211.725	&	1.21E+07	&2.563 	&	4.87E-02	 	&	6.68E-02	 	\\
	&	5p$^3$ $^2$D$^o_{5/2}$	&	217.919	&	1.67E+08	 &0.988	&	7.16E-01	 	&	9.27E-01	 	\\
	&		&		&				&				&				\\
5d $^4$P$_{3/2}$	&	5p$^3$ $^4$S$^o_{3/2}$	&	178.089	&	4.40E+07&1.375	 &	8.36E-02	 	&	2.34E-01	 	\\
	&	5p$^3$ $^2$D$^o_{3/2}$	&	209.842	&	1.11E+08	 	&1.037&	2.93E-01 	&	5.90E-01	 	\\
	&	5p$^3$ $^2$D$^o_{5/2}$	&	215.925	&	5.08E+06 	&1.248&	1.42E-02	 	&	2.70E-02	 	\\
	&	5p$^3$ $^2$P$^o_{1/2}$	&	251.456	&	4.76E+06	 	&1.063&	1.81E-02	 	&	2.53E-02	 	\\
	&	5p$^3$ $^2$P$^o_{3/2}$	&	265.260	&	2.33E+07	 	&0.673&	9.83E-02	 	&	1.24E-01	 	\\
	&		&		&				&				&				\\
5d $^4$F$_{7/2}$	&	5p$^3$ $^2$D$^o_{5/2}$	&	214.184	&	1.63E+07&0.982	 	&	9.00E-02	 	&	9.94E-01	 	\\
	&		&		&				&				&				\\
5d $^4$P$_{1/2}$	&	5p$^3$ $^4$S$^o_{3/2}$	&	176.309	&	6.17E+06*	&0.081* 	&	5.75E-03*	 	&	3.03E-02*	 	\\
	&	5p$^3$ $^2$P$^o_{1/2}$	&	248.044	&	5.24E+06*	 	&0.840*&	9.68E-03*	 	&	2.57E-02*	 	\\
	&	5p$^3$ $^2$P$^o_{3/2}$	&	261.466	&	1.85E+08*	 	&13.603*&	3.79E-01*	 	&	9.07E-01*	 	\\
	&		&		&				&				&				\\
5d $^2$P$_{3/2}$	&	5p$^3$ $^4$S$^o_{3/2}$	&	176.202	&	1.84E+07	&1.920 	&	3.43E-02	 	&	1.57E-01	 	\\
	&	5p$^3$ $^2$D$^o_{3/2}$	&	207.312	&	1.94E+06	 	&0.576&	5.01E-03	 	&	1.65E-02	 	\\
	&	5p$^3$ $^2$P$^o_{1/2}$	&	247.832	&	2.47E+07	 &1.016	&	9.11E-02	 	&	2.10E-01	 	\\
	&	5p$^3$ $^2$P$^o_{3/2}$	&	261.230	&	7.09E+07	 &1.103&	2.90E-01	 	&	6.04E-01	 	\\
	&		&		&				&				&				\\
5d $^2$F$_{5/2}$	&	5p$^3$ $^4$S$^o_{3/2}$	&	174.560	&	1.08E+06	 &0.400	&	2.96E-03	 	&	1.07E-02	 	\\
	&	5p$^3$ $^2$D$^o_{3/2}$	&	204.958	&	9.93E+07	 	&0.919&	3.75E-01	 	&	9.81E-01	 	\\
	&		&		&				&				&				\\

\end{tabular}
\end{ruledtabular}
\end{table*}

\addtocounter{table}{-1}

\begin{table*}
\caption{ Continued.}
\begin{scriptsize}
\begin{ruledtabular}
\begin{tabular}{llrcccc}
\multicolumn{1}{c}{Upper Level$^a$}&\multicolumn{1}{c}{Lower Level$^a$}&
\multicolumn{1}{c}{$\lambda$(nm)$^a$}&
\multicolumn{1}{c}{$A_{ki}$ (s$^{-1}$)$^b$} &
\multicolumn{1}{c}{r}&
\multicolumn{1}{c}{$gf_{ik}^b$} &
\multicolumn{1}{c}{$BF^b$}\\
\hline

5d $^4$D$_{1/2}$	&	5p$^3$ $^4$S$^o_{3/2}$	&	173.557	&	2.74E+06*	&0.197* 	&	2.48E-03*	 	&	1.78E-02*	 	\\
	&	5p$^3$ $^2$P$^o_{1/2}$	&	242.634	&	2.05E+07*	 	&0.949*&	3.62E-02*	 	&	1.33E-01*	 	\\
	&	5p$^3$ $^2$P$^o_{3/2}$	&	255.462	&	1.27E+08*	 	&2275.0*&	2.48E-01* 	&	8.21E-01*	 	\\
	&		&		&				&				&				\\
5d $^4$D$_{5/2}$	&	5p$^3$ $^4$S$^o_{3/2}$	&	169.823	&	3.39E+06*	&0.521* 	&	8.81E-03*	 	&	2.13E-02*	 	\\
	&	5p$^3$ $^2$D$^o_{3/2}$	&	198.542	&	5.57E+07*	 	&2.040*&	1.98E-01*	 	&	3.50E-01*	 	\\
	&	5p$^3$ $^2$D$^o_{5/2}$	&	203.979	&	3.86E+07*	 &1.191*&	1.45E-01*	 	&	2.42E-01*	 	\\
	&	5p$^3$ $^2$P$^o_{3/2}$	&	247.458	&	6.09E+07*	&10.130* 	&	3.36E-01*	 	&	3.82E-01*	 	\\
	&		&		&				&				&				\\
5d $^2$F$_{7/2}$	&	5p$^3$ $^2$D$^o_{5/2}$	&	195.039	&	2.60E+08	&1.070 	&	1.18E+00	 	&	9.97E-01	 	\\
\end{tabular}
\end{ruledtabular}
\end{scriptsize}
\begin{itemize}
\item[$^a$] Hassini {\it et al.} \cite{HARVW88}. The wavelengths are determined from the experimental
energy levels and are given in air when longer than 200 nm.
\item[$^b$] MCDHF calculation (this work).
The MCDHF values are corrected from the experimental transition energies.
Only the Babushkin gauge results are quoted.
\item[*] Affected by convergence problems (see the text).
\end{itemize}
\end{table*}

\end{document}